\documentclass[aps,prb,superscriptaddress,showpacs,preprint]{revtex4-1}
\usepackage{graphicx} 
\usepackage[margin=1.0 in]{geometry}
\usepackage{lipsum}
\usepackage{dcolumn}
\usepackage{bm}
\usepackage{xcolor}
\usepackage{hyperref}

\begin{document}

\title{Nuclear Quantum Effects on the Electronic Structure of Water and Ice}
  
\author{Margaret Berrens}
\affiliation
{Department of Chemistry, University of California Davis, One Shields Ave. Davis, CA, 95616.}
\author{Arpan Kundu}
\affiliation
{Pritzker School of Molecular Engineering, University of Chicago, Chicago, IL 60637}
\author{Marcos F. Calegari Andrade }
\affiliation
{Quantum Simulations Group, Materials Science Division, Lawrence Livermore National Laboratory, Livermore, California, 94550-5507}
\author{Tuan Anh Pham}
\affiliation
{Quantum Simulations Group, Materials Science Division, Lawrence Livermore National Laboratory, Livermore, California, 94550-5507}
\author{Giulia Galli}
\affiliation
{Pritzker School of Molecular Engineering, University of Chicago, Chicago, IL 60637}
\affiliation{Department of Chemistry, University of Chicago, Chicago, Illinois 60637, United States}
\affiliation{Materials Science Division and Center for Molecular Engineering, Argonne National Laboratory, Lemont, Illinois 60439, United States}
\author{Davide Donadio}
\email{ddonadio@ucdavis.edu}
\affiliation
{Department of Chemistry, University of California Davis, One Shields Ave. Davis, CA, 95616.}

\begin{abstract}

The electronic properties and optical response of ice and water are intricately shaped by their molecular structure, including the quantum mechanical nature of hydrogen atoms. In spite of numerous studies appeared over decades, a comprehensive understanding of the effect of the nuclear quantum motion on the electronic structure of water and ice at finite temperatures remains elusive.
Here, we utilize molecular simulations that harness the efficiency of machine-learning potentials and many-body perturbation theory to assess the impact of nuclear quantum effects on the electronic structure of water and hexagonal ice. By comparing the results of path-integral and classical simulations, we find that including nuclear quantum effects leads to a larger renormalization of the fundamental gap of ice, compared to that of water, eventually leading to a comparable gap in the two systems, consistent with experimental estimates. Our calculations suggest that the quantum fluctuations responsible for an increased delocalization of protons in ice, relative to water,  are a key factor leading to the enhancement of nuclear quantum effects on the electronic structure of ice. 

\end{abstract}
\maketitle

Water and ice provide a natural solvation environment for most chemical processes of atmospheric and biological importance. 
The electronic structure of water affects the chemistry of biological, geochemical, and environmental reactions, as well as the properties of energy conversion devices, such as photoelectrochemical cells.\cite{zhong_atmospheric_2019, sverjensky_water_2014, walter_solar_2010, gratzel_photoelectrochemical_2001} 
Understanding the relation between the structural and electronic properties of water will help shed light on the chemistry of natural processes and aid the optimization of technological applications. 
There has been extensive research on the electronic properties of liquid water\cite{pham_2014, chen_ab_2016,  gaiduk_electron_2018, bischoff_band_2021, Tal_2024}, but significantly less attention has been directed towards the electronic structure of hexagonal ice (Ice I$_h$), water's most common solid form. 
A thorough understanding of its electronic structure is essential to elucidate the role of ice as a substrate or a catalyst in environmental chemistry,\cite{grannas_role_2014, klanova_environmental_2003, takenaka_chemical_2007, grannas_overview_2007} and astrochemistry.\cite{mifsud_sulfur_2021, jenniskens_structural_1994} 

Data inferred from experiments and interpreted by Chen et al. point at water and ice having similar fundamental gaps ($E_g$),  despite considerable structural differences: $9.0\pm 0.2$ for water\cite{bernas_electronic_1997, winter_full_2004, ambrosio_2017, gaiduk_electron_2018} at room temperature and $9.4\pm 0.3$ for ice\cite{campbell_1979, shibaguchi_1977, baron_1978} at 77 K. 
Interpreting experiments at the molecular level and connecting the structural and dynamical properties to the electronic structure of water and ice remain challenging tasks, especially due to the disordered and dynamic nature of these systems. Molecular dynamics (MD) simulations combined with electronic structure calculations offer a promising means to address this knowledge gap.\cite{gaiduk_electron_2018} Whereas nuclei are treated as classical particles in first-principles MD (FPMD), the inclusion of nuclear quantum effects (NQE) is necessary to describe the electron-phonon interactions affecting the properties of materials with light elements.\cite{ceriotti_nuclear_2016, markland_2018, Kundu_PRM_2021,kundu_influence_2022,Alvertis_PRB_2022, Kundu_JCTC_2023,Kundu_NV_JPCL_2024} 
Nuclear quantum dynamics can be modeled using Feynman's path integrals (PI) or related methods,\cite{PI_Berne_Rev_1986,PI_Marx_Rev_1996,PI_hererro_rev_2014,QT_Review_Finocchi_2022} although at a substantially higher computational cost. 

Using trajectories generated with  the Vydrov and Van Voorhis (rVV10)
van der Waals density functional,\cite{sabatini_2013} path-integral MD (PIMD) simulations with 6 beads, and supercells with 32 water molecules, Chen {\it et al.} estimated the bandgap of water at 300~K using self-consistent GW calculations with a two-point exchange-correlation kernel.\cite{chen_ab_2016} They obtained a band gap of 9.8 and 8.9 eV without and with vertex corrections, respectively, and in both cases a renormalization ($\Delta E_g^{NQE}$) of -0.7 eV. A similar computational protocol was used by Tal et al. \cite{Tal_2024} who however increased the number of frames over which they conducted MBPT calculations and slightly modified the treatment of the exchange-correlation Kernel, obtaining a band gap of 9.2 eV when using vertex corrections. Gaiduk {\it et al.} reported a bandgap of  10~eV with a $\Delta E_g^{NQE}$ of -0.5 eV, using G$_0$W$_0$ calculations starting from hybrid DFT (they used both range-separated (RSH) and self-consistent hybrids)\cite{gaiduk_electron_2018}. They conducted simulations with the MB-pol potential, PIMD with 32 beads and used 64 water molecule cells. The major difference between the results obtained by Tal {et al.} \cite{Tal_2024} and Gaiduk {\it et al.} originates from the different positions of the conduction band predicted in the two papers, while the position of the valence band is similar. 
For ice I$_h$,  Engel et al. and Monserrat et al. estimated a large energy gap renormalization, -1.5 eV, using stochastic methods based on the quasi-harmonic approximation\cite{engel_vibrational_2015, monserrat_giant_2015}, though the accuracy of these methods in a strong anharmonic regime is not guaranteed.\cite{Alvertis_PRB_2022, Kundu_JCTC_2023}. Bischoff et al. reported a band gap of ice of 9.8 and 9.3 eV (and that of water of 9.6 and 9.1 eV), depending on the type of self-consistent GW method used. In the case of ice, they applied just the zero-point renormalization (ZPR) obtained in Ref.\cite{engel_vibrational_2015, monserrat_giant_2015} to electronic structure calculations at zero T, carried out for a Bernal-Fowler unit cell with 12 molecules.\cite{bischoff_band_2021}
The origin of the difference between the impact of NQE on the electronic properties of water and ice has not yet been clarified, nor has $\Delta E_g^{NQE}$ been computed for ice by MD at finite temperatures. In addition, comparisons have been made for calculations carried out at different levels of electronic structure theory and with different structural models (different force fields and/or density functionals). 
Clearly, a fair and robust comparison calls for modeling the dynamical properties of water and ice at finite temperatures on an equal footing, using quantum simulations and computing the electronic structure of both systems within the same approximations. Such a comparison is still amiss.  

In this letter, we investigate the effect of NQEs on the electronic structure of water and ice I$_h$ by carrying out MBPT calculations at the G$_0$W$_0$ level of theory for an ensemble of configurations obtained from classical and PIMD simulations, to provide a consistent estimate of the bandgap renormalization of both systems.
Our calculations confirm that NQEs have a significantly larger effect on the electronic properties of ice $I_h$ compared to those of liquid water. A comparative analysis of hydrogen-bonding configurations in water and ice suggests that the degree of transient proton transfer from hydrogen bond donors to acceptors is the key structural feature affecting the band-gap renormalization.    


We use machine learning potentials (MLP) to accelerate both classical and PIMD simulations. We specifically employ two different MLPs derived from density functional theory (DFT) with two functionals: a van der Waals corrected hybrid functional (revPBE0-D3),\cite{perdew_rationale_1996,adamo_toward_1999,grimme_consistent_2010} and the {\it strongly constrained and appropriately normed} (SCAN) exchange-correlation functional.\cite{sun_strongly_2015} 
Using revPBE0-D3, we obtained a neural network potential fitted with an evolutionary algorithm; this potential has been shown to reproduce accurately the thermodynamic properties of water and ice, including the density anomaly of water\cite{fan_neuroevolution_2021,chen_thermodynamics_2024}, and   
 it is of similar quality as previously reported neural network models fitted on the same density functional. However it is more computationally efficient.\cite{cheng_2019,schran_committee_2020}
In the case of the SCAN functional, we used a previously trained and validated deep neural network potential (DNNP) that accurately reproduces structural \cite{zhang_phase_2021} and vibrational\cite{msommers_raman_2020}  properties of water and ice.
We compared electronic structure results obtained for trajectories with two different MLPs to rule out systematic biases originating from a specific MLP scheme or the underlying density functional approach. 
The differences and similarities of the two underlying density functionals used in this work have been the focus of several \textit{ab intio} MD studies and are briefly discussed in the Supporting Information.\cite{ohto_2019, Palos_2022} 
Our results are further validated using trajectories generated with a DNNP fitted to the first-principles many-body force field MB-pol.\cite{bore_2023}

We performed quantum simulations with a generalized Langevin equation thermostat (PIGLET),\cite{PIGLET_Ceriotti_PRL_2012, kapil_i-pi_2019} which provides well-converged results, compared to PIMD, using a small number of beads ($p=8$) in the ring-polymer.\cite{Kapil_JCP_2016} MD simulations were carried out at 300 K and 230 K for water and ice respectively. A temperature of 230 K was chosen for ice to investigate the NQE on electronic properties at environmentally relevant conditions. It is expected that as the temperature increases the gap will only slightly decrease for Ice $I_h$.\cite{engel_vibrational_2016} For the PIGLET simulations, at 230 K, the average quantum kinetic and potential energy of ice is converged within 3~meV/atom with only 8 beads (Figure S2). 
Size effects on the calculated bandgap of liquid water were carried out by~\citet{gaiduk_electron_2018} Here, we tested size effects on the bandgap of bulk ice using supercells with 96 and 192 molecules, which showed that a simulation cell of 96 water molecules is sufficient to obtain an estimate of the bandgap within the statistical uncertainty of our calculations (Table S1). The choice of system size in our MD simulations was mainly constrained by the cost of performing several tens of MBPT calculations over classical MD and PIMD trajectories.

Electronic structure calculations are performed using the G$_0$W$_0$ method on electronic structure calculations at the PBE \cite{perdew_generalized_1996} level of theory, for 50 equally spaced frames selected from MD and  PIMD trajectories of 100 ps for water and ice, respectively (convergence as a function of the number of frames is shown in Figure S3). The electronic structure calculations from the PIMD runs were performed for a randomly chosen ring-polymer bead. All G$_0$W$_0$ calculations were carried out with the WEST code.\cite{govoni_large_2015}
We note that G$_0$W$_0$@PBE underestimates the $E_g$ of water, i.e. the energy difference between the conduction band minimum (CBM) and valence band maximum (VBM), of liquid water by approximately 1~eV lower, compared to higher level theories using hybrid functionals as starting points for MBPT and compared to experimental estimates\cite{gaiduk_electron_2018, pham_2014}.  The use of the quasiparticle self-consistent GW (QSGW) method that includes vertex corrections also yields larger gaps\cite{chen_ab_2016, bischoff_band_2021}. However, for liquid water one-shot G$_0$W$_0$@PBE calculations yield an estimate of the effect of NQE on the electronic structure which is similar to that of the  QSGW method.\cite{chen_ab_2016}. Since the focus of this work is to estimate NQEs on the gap rather than the absolute values of the $E_g$, we have utilized  G$_0$W$_0$@PBE, which allowed us to sample a well-converged ensemble of configurations for relatively large systems.


\begin{figure}[h!]
    \centering
    \includegraphics[width=0.75\linewidth]{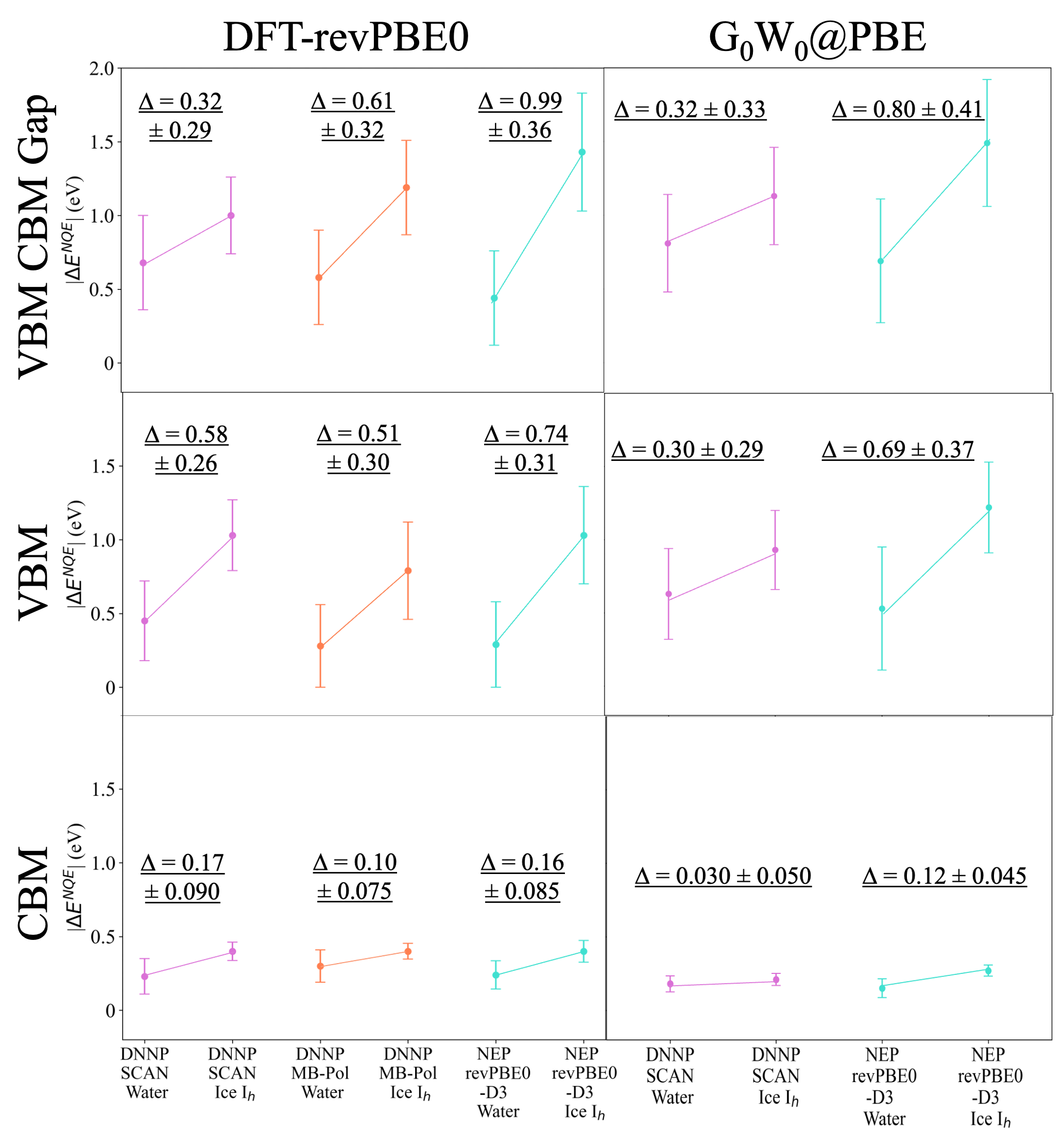}
    \caption{The difference between the fundamental gap obtained with classical and quantum trajectories ($\Delta E_g^{NQE}$), upper panel, and the difference between the position of the VBM (middle panel) and CBM (lower panel) obtained with classical and quantum trajectories,  for liquid water and Ice $I_h$. The results on the left panels were obtained  at the DFT level using the revPBE0 functional for 100 frames (left) and those on the right panels at the G$_0$W$_0$ level for 50 frames (right). The x-axis indicates the Ml potentials used in the calculations. $\Delta$ represents the difference in $\Delta E_g^{NQE}$ between water and ice for each ML potential.}
    \label{fig:barplot}
\end{figure} 

The difference between the $E_g$ of water and ice obtained with classical and quantum simulations, as well as the difference between the positions of the VBM, and CBM, obtained from G$_0$W$_0$@PBE and hybrid calculations, are shown in Figure~\ref{fig:barplot}. 
Regardless of the underlying potential used for sampling and the electronic structure method, Figure~\ref{fig:barplot} shows that NQEs considerably reduce the bandgap of both water and ice by raising the energy of the VBM and lowering, to a much smaller extent, the energy of the CBM. Consistent with previous findings, we find that NQEs have a much larger impact in ice, for which the $\Delta E_g^{NQE}$ amounts to 1.1 $\pm$ 0.32 -- 1.5 $\pm$ 0.43 eV on the band gap, which instead decreases by 0.70 $\pm$ 42 -- 0.81 eV $\pm$ 0.33 in water. Hence, the larger NQEs in ice overall lead to a similar $E_g$ for ice and liquid water, consistent with data inferred from experiments ($9.0\pm 0.2$ eV for water and $9.4\pm 0.3$ eV for ice).
Our results for  $\Delta E_g^{NQE}$ are consistent with previous calculations for both water (-0.7 eV in \cite{chen_ab_2016}  and -0.5 in \cite{gaiduk_electron_2018}) and ice (-1.52 eV in \cite{engel_vibrational_2015}). 
We note that the value for ice I$_h$ in Ref.~\cite{engel_vibrational_2015} was obtained for a configuration optimized at zero temperature. We find a similar bandgap renormalization of 1.65 eV due to NQEs using a geometry-optimized structure. 

Overall, sampling with NEP enhances the difference in the NQEs-induced bandgap renormalization between ice and water, although it remains within the same uncertainty as the DNNP results.  
The error bars on the $\Delta E^{NQE}$ for the VBM are much larger than for the CBM. The main effect of quantum dynamics is not only to raise the energy of the VBM state, especially in ice, but also to broaden its energy distribution. In turn, the CBM is only slightly affected and does not undergo significant broadening. 

Using  the hybrid revPBE0 functional \cite{perdew_rationale_1996,adamo_toward_1999,grimme_consistent_2010}, we  verified that the trends reported above for $\Delta E^{NQE}$ for water and ice hold for trajectories generated with MB-pol data.\cite{bore_2023, Medders_2014, babin_2014, babin_2013} This model provides an accurate description of the structural, thermodynamic, dynamical, and spectroscopic properties of water\cite{cole_2016, moberg_2017, medders_2016}\cite{bore_2023}. Finally, the corresponding $\Delta E_g^{NQE}$'s computed from the Kohn-Sham orbitals using the SCAN functional are reported in Figure S5. The values of the VBM and CBM energies from the MBPT and the Kohn-Sham DFT calculations are also reported in Tables S2, S3, and S4, respectively. We find that inexpensive DFT calculations (such as those using the SCAN functional) reproduce the trends for NQEs obtained from accurate MBPT, but, unsurprisingly, MBPT is necessary to achieve bandgaps in reasonable agreement with experiments. 
As expected, the use of a hybrid exchange and correlation functional brings DFT calculations much closer to MBPT results and results for $\Delta E_g^{NQE}$ are within statistical uncertainty. The underestimation of the bandgap with revPBE0 can be in part attributed to the fact that the inverse of the appropriate fraction of exact exchange ($\alpha$) in the hybrid functional is more than two times larger than that of the electronic dielectric constant ($\varepsilon_\infty$) of water.\cite{pham_2014}


\begin{figure}[t]
    \centering
    \includegraphics[width=0.7\linewidth]{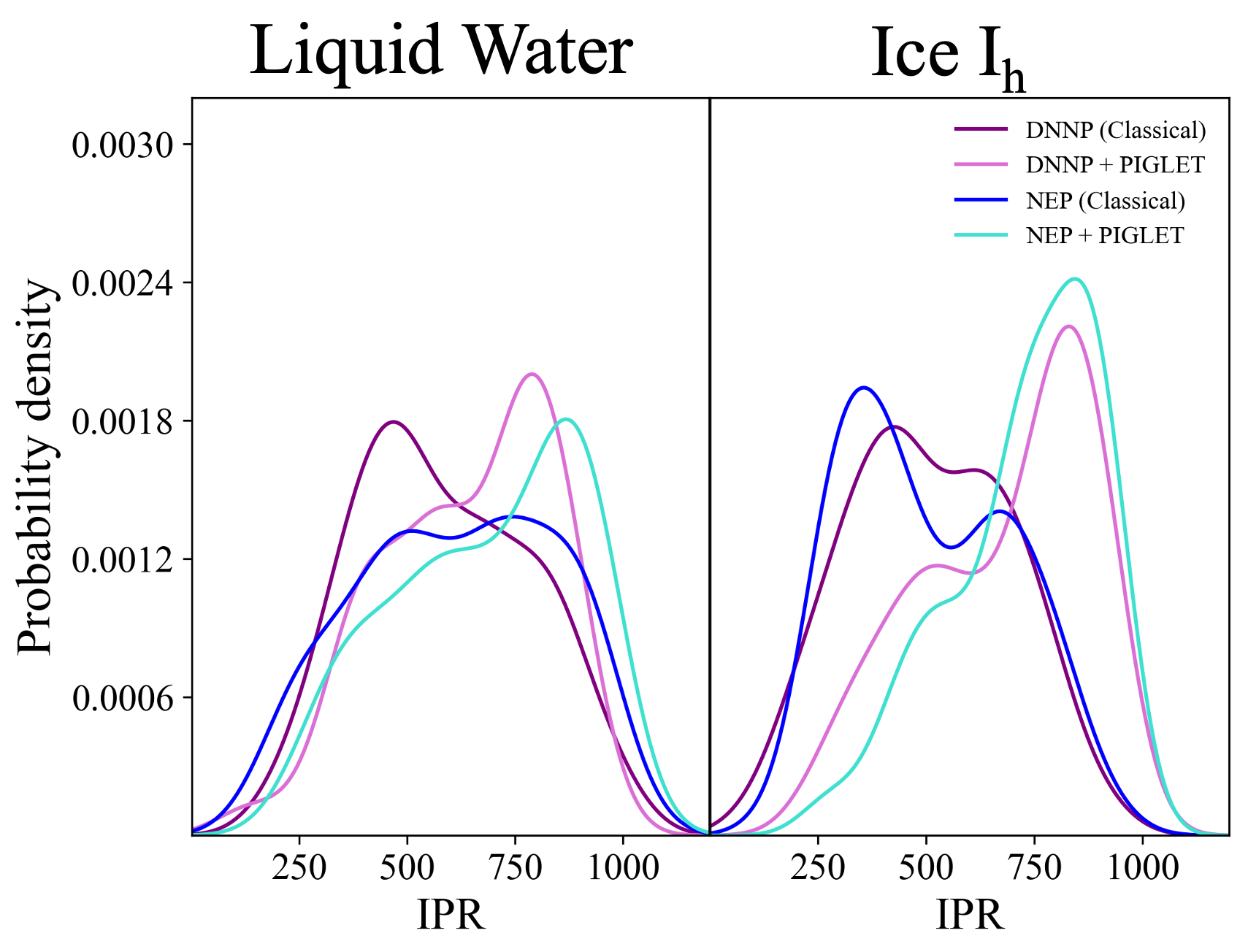}
    \caption{Probability density of the inverse participation ratio (IPR), fit with a Gaussian kernel density estimation, for the VB calculated with SCAN eigenstates, for liquid water (left) and ice (right). For each water system, we report the IPR for the DNNP with (purple) and without (turquoise) NQEs, and for the NEP with (light purple) and without (green) NQEs. A higher IPR indicates stronger localization. The IPR values are scaled by 10000.}
    \label{fig:ipr}
\end{figure}

To understand the significant effect of quantum dynamics on the position of the VBM we calculated the inverse participation ratio (IPR) for an ensemble of snapshots from either PIMD or classical MD simulations (Figure~\ref{fig:ipr}). The IPR of the $i$th Kohn-Sham orbital is calculated as $ \int |\psi_i|^4 d^3r/(\int|\psi_i|^2 d^3r)^2 $ where a higher IPR value corresponds to a more localized single-particle wave function. We calculated IPRs for the Kohn-Sham orbitals using the SCAN functional including states within 50 meV of the VBM, which we refer to as the valence band (VB). For each water system and potential, we calculated the Gaussian kernel density estimation of the IPR for the classical and quantum simulations. The histogram of IPR values for both the VB and the conduction band (CB) is reported in Figure S6.
The IPR distribution for the VB spans values over two orders of magnitude larger than those for the CB, consistent with the known localized nature of occupied levels as opposed to delocalized empty levels. For both water and ice, the NQEs shift the VB distribution toward higher IPR values, indicating that quantum delocalization of the nuclei enhances the localization of the electronic levels. This phenomenon, previously observed in liquid water by Chen et al.,\cite{chen_ab_2016} was attributed to Anderson localization resulting from an increased disorder in the proton distributions, due to quantum fluctuations. The increased localization of electronic levels, akin to surface effects observed in water,\cite{gaiduk_electron_2018} may also stem from the weakening of hydrogen bonds due to quantum delocalization.
Since water and ice have different cell sizes, we cannot compare the absolute shift of the IPR as IPR depends on cell size. Therefore we report relative IPR shift  with respect to the position of the peak of IPR distribution in the classical simulation. For water and ice, there is an average shift in the relative peak position of the IPR distribution of the VB of 18\% and 66\%, respectively. The more pronounced shift for the VB of ice I$_h$ is responsible for the larger bandgap renormalization.

\begin{figure}[t]
    \centering
    \includegraphics[width=1\linewidth]{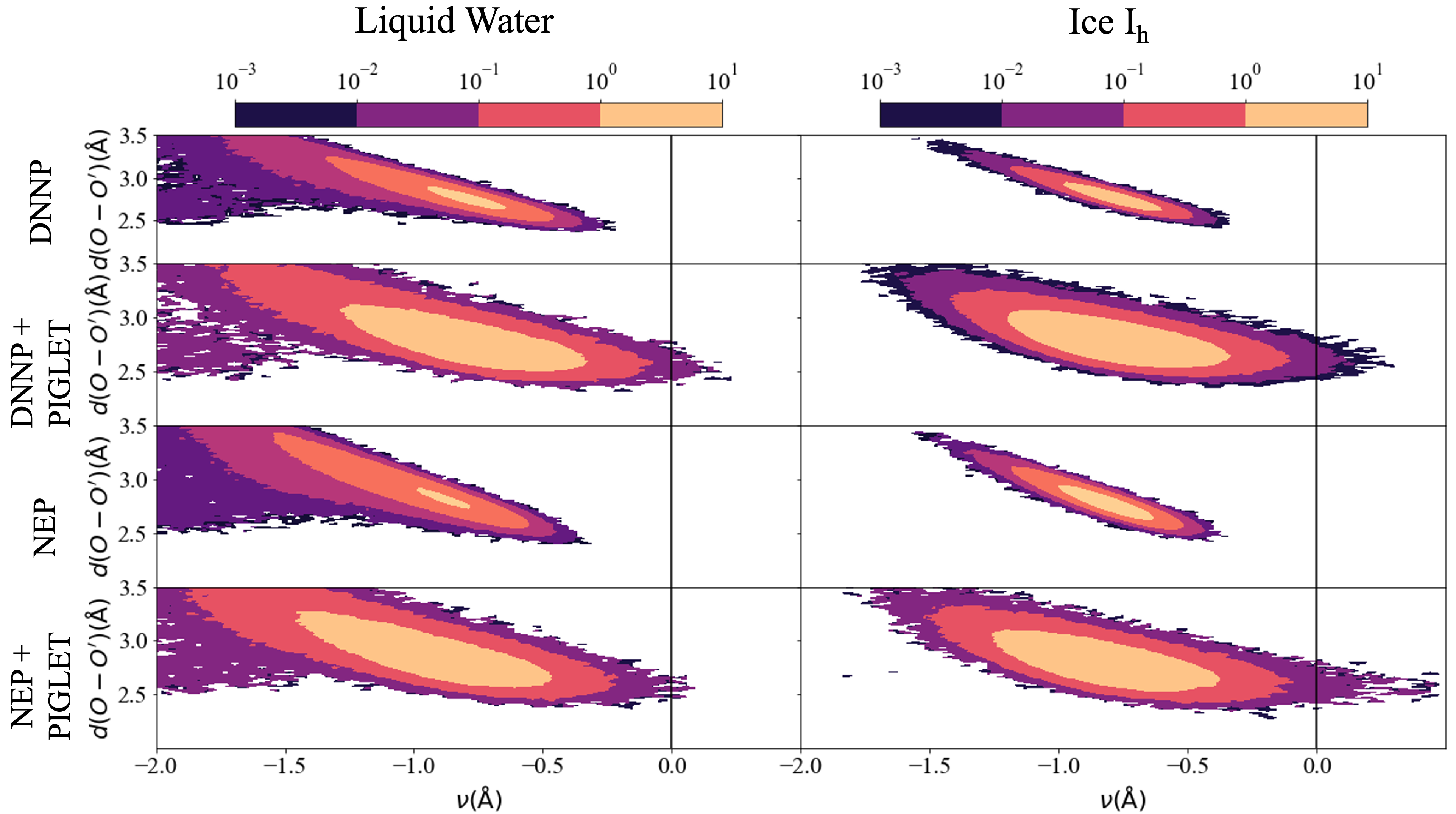}
    \caption{Joint probability distribution of the proton-transfer coordinate $\nu$ and the distance d(O–O') between the covalently bound and acceptor oxygen atoms for liquid water (left) and Ice $I_h$ (right). The top two rows are a comparison of the sampling with DNNP with and without NQEs, and the bottom two rows are a comparison of the sampling with NEP with and without NQEs.}
    \label{fig:protontransfer}
\end{figure}
Previous studies investigating the NQEs on the structural properties of liquid water found that the inclusion of NQEs softens the local structure of water due to proton delocalization.\cite{markland_2018} In particular, the oxygen-hydrogen and hydrogen-hydrogen radial distribution functions (g$_{OH}$ and g$_{HH}$) exhibit the most apparent structural impatc from NQEs. Specifically, the NQEs cause a broadening of the first two peaks of g$_{OH}$, which suggests a higher probability of proton transfer.
The calculated radial and angular distribution functions in Figure S7 confirm that both MLPs reproduce these structural features for both water and ice. Similarly, NQEs cause a broadening of the distribution function of the H-O-H angle for water and ice. However, the causes of the difference in the NQE-induced band-gap renormalization between water and ice, cannot be inferred from differences in radial distribution functions. 
To investigate why $\Delta E_g^{NQE}$ is much larger in ice I$_h$ than in liquid water, we calculated the distributions of proton transfer coordinates and oxygen-oxygen distances for each hydrogen-bonded pair of water molecules (Figure~\ref{fig:protontransfer}) in our simulation cells.
The proton transfer coordinate is defined as $\nu = d_{OH} - d_{O'H}$, where $d_{OH}$ is the distance between O and H in the covalent bond and $d_{O'H}$ is the distance between the same hydrogen atom and the acceptor hydrogen-bonded oxygen $O^\prime$.\cite{ceriotti_nuclear_2013,fritsch_nuclear_2014}  
Incorporating NQEs in the simulations enhances hydrogen delocalization, yielding a small fraction of configurations with $\nu > 0$ that corresponds to short-lived autoprotolysis events. Here, for liquid water, we see that NQEs broaden the joint probability distribution $P(d_{OO'},\nu)$ with a tail extending into the $\nu > 0$ region. Comparing the distributions obtained with different MLPs and PIGLET, we observe a two times larger fraction of hydrogen bond configurations with $\nu > 0$ with the DNNP. 
In classical MD simulations, the probability distribution $P(d_{OO'},\nu)$ for ice I$_h$ is much narrower compared to that of water. However, the inclusion of NQEs results in a broadening of the $P(d{OO'},\nu)$ distribution for ice, causing it to resemble that of water. Notably, this broadened distribution exhibits a higher likelihood of transient autoprotolysis in ice than in water after accounting for the NQEs. 

\begin{figure}[t]
    \centering
    \includegraphics[width=1\linewidth]{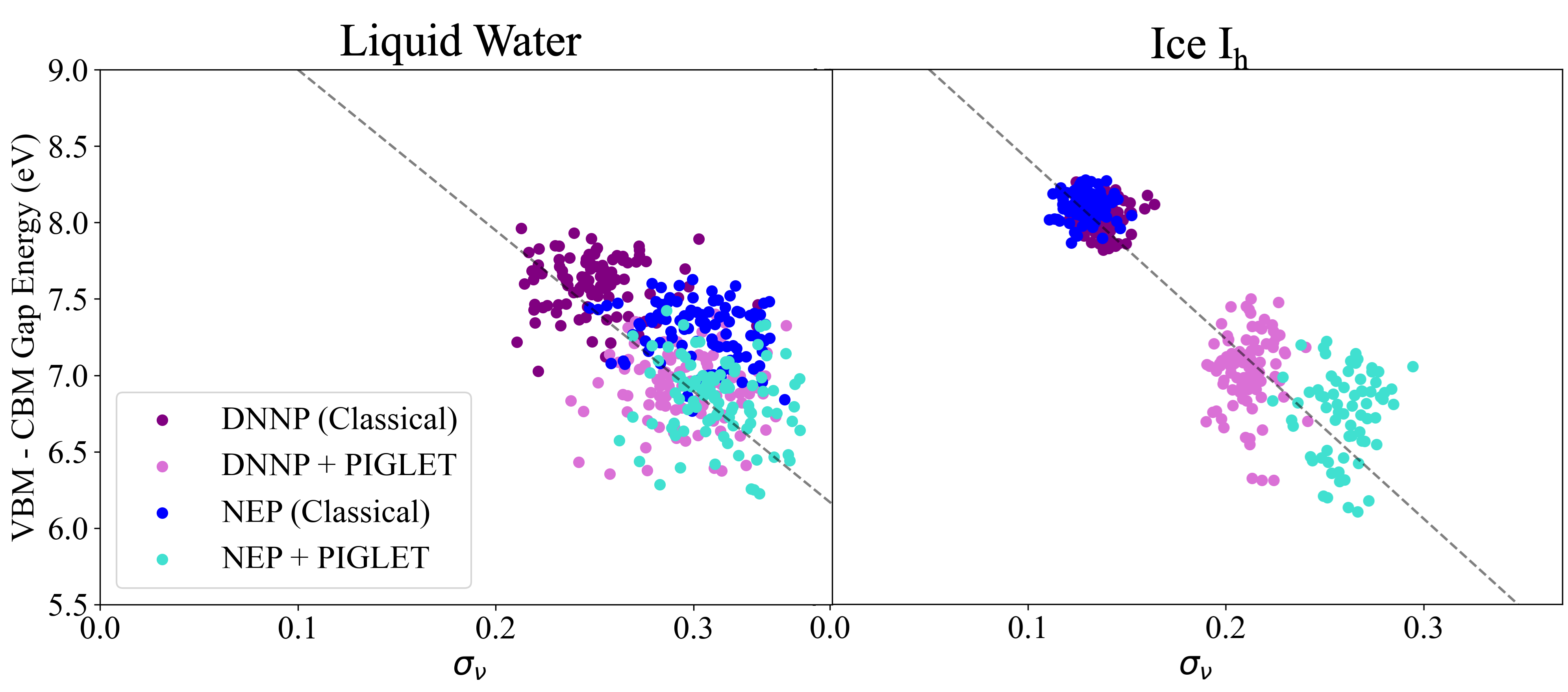}
    \caption{Scatter plot of the standard deviation in the proton transfer coordinate ($\sigma_{\nu}$) for all hydrogen bonds in a given water system versus the corresponding VBM-CBM bandgap for a given frame calculated at the DFT level (revPBE0 functional) for liquid water (left) and Ice $I_h$ (right).}
    \label{fig:correlation}
\end{figure} 



To verify whether the observed change in the structure of the hydrogen bonds is responsible for the NQEs on E$_g$, we plot E$_g$ (calculated at the DFT level with the revPBE0 functional) against the standard deviation of the proton transfer coordinate $(\sigma_\nu)$ for each snapshot; this coordinate is a measure of the spatial fluctuations in the proton transfer coordinates. Our results for classical and quantum simulations for both water and ice show that there is a correlation between the value of E$_g$ and the proton fluctuations in the hydrogen bonds of the system, with a larger $\sigma_\nu$ associated with a lower band-gap.

When comparing water and ice, it is clear that for each system, a larger value of $\sigma_\nu$ correlates with lowered bandgaps. 
Significant distinctions in the distributions of $\sigma_\nu$ emerge between classical and quantum simulations for ice, with pronounced variations accentuated when employing the NEP potential, which yields larger NQEs on the bandgap. 
In contrast, the distributions for liquid water exhibit not only considerably smaller disparities between classical and quantum simulations but also similar trends for both MLPs potential energy landscapes. 
These findings confirm that there exists a correlation between the fluctuations of protons in the hydrogen bond and the influence of NQEs on the bandgap. 
Previous investigations on liquid water \cite{ceriotti_nuclear_2013} have also reported that NQEs induce substantial rearrangements in the positions of Wannier centers along a proton transfer coordinate, underscoring the pivotal role of proton delocalization in shaping the electronic structure of liquid water. Figure~\ref{fig:correlation}  highlights that NQEs exert an even more substantial impact on hydrogen bond fluctuations in Ice $I_h$, consistent with their greater influence on the bandgap renormalization.

In summary, we have calculated the finite-temperature bandgap renormalization of water and ice I$_h$ using many-body perturbation theory on ensembles of configurations generated using either classical MD or PIMD simulations, and employing two different machine-learned potentials fitted to DFT forces obtained with meta-GGA or hybrid functionals. 
Consistent with earlier studies,\cite{bischoff_band_2021} we find that nuclear quantum effects induce a significant decrease in the electronic bandgap of water, and in ice, the reduction is even more pronounced, reaching twice the magnitude observed in water. As a consequence water and Ice $I_h$ turn out to have similar fundamental gaps, consistent with data inferred from experiments.
The two sets of simulations with different machine-learned potentials exhibit the same trends on nuclear quantum effects, albeit with a few quantitative differences, stemming from slight distinctions in the structural features of hydrogen bonds in classical simulations with the two potentials. 
A statistical analysis of the differences in the local structure of hydrogen bonds in quantum and classical simulations suggests that proton delocalization, and the resulting enhanced transient autoprotolysis, observed in PIMD simulations are the main cause for the much larger nuclear quantum effects on the bandgap renormalization of ice I$_h$, compared to water. 
These results highlight the critical importance of accounting for nuclear quantum effects when modeling the electronic properties of water, and even more so of ice I$_h$.
The observed influence of quantum proton delocalization and the ensuing bandgap renormalization are poised to have a considerable impact on molecular simulation studies examining ice as a solvation medium for environmental pollutants and as a catalyst for atmospheric chemistry reactions within snowpacks or clouds, a topic that has garnered increasing interest.\cite{bartels-rausch_review_2014, bononi_bathochromic_2020, berrens_solvation_2023, loerting_2006}


\section*{Conflicts of interest}
There are no conflicts to declare.

\section*{Supporting Information}
Computational Methods, discussion of differences between MLPs, vibrational density of states, convergence of number of beads and number of frames used for electronic structure calculations, discussion of sizes effects on the proton transfer coordinate and bandgap values, a summary of calculated VBM CBM gap values, radial and angular distribution functions, histograms of IPR, and electronic density of states. 

\section*{Data availability}
\textcolor{black}{Data are available in the Materials Cloud Archive (www.materialscloud.org) with ID materialscloud:xxxx}

\section*{Acknowledgements}
We are grateful to Jiawei Zhan and Wenzhe Yu for their assistance with the G$_0$W$_0$ calculations, and to Yifan Li for his assistance in the proton transfer coordinate calculations. We thank Francesco Paesani for useful suggestions about the choice of ML potentials. is This work partly supported by the National Science Foundation under Grants No. 2305164 and No. 2053235, and by the Advanced Materials for Energy-Water Systems Center, an Energy Frontier Research Center funded by the U.S. Department of Energy, Office of Science, Basic Energy Sciences.  
Many body perturbation theory calculations were carried out with the WEST code, whose development is supported by MICCoM, as part of the Computational Materials Sciences Program funded by the U.S. Department of Energy, Office of Science, Basic Energy Sciences, Materials Sciences, and Engineering Division through Argonne National Laboratory, under Contract No. DE-AC02-06CH11357. 
G$_0$W$_0$ calculations have been performed using resources of the Argonne Leadership Computing Facility, which is a DOE Office of Science User Facility, supported by an ALCC grant. The work at the Lawrence Livermore National Laboratory was performed under the auspices of the U.S. Department of Energy under contract DE-AC52-07NA27344. M.F.C.A. and T.A.P. were supported as part of the Center for Enhanced Nanofluidic Transport, an Energy Frontier Research Center funded by the U.S. Department of Energy, Office of Science, Basic Energy Sciences under award DE-SC0019112.

\bibliography{elecstruc}

\end{document}